**Triplet Exciton Sensitization of Silicon Mediated by Defect States in Hafnium Oxynitride**

*Narumi Nagaya\*, Alexandra Alexiu, Collin F. Perkinson, Oliver M. Nix, Moungi G. Bawendi, William A. Tisdale, Troy Van Voorhis, Marc A. Baldo*

N. Nagaya, C. F. Perkinson, O. M. Nix, M. A. Baldo

Research Laboratory of Electronics, Massachusetts Institute of Technology, 77 Massachusetts Avenue, Cambridge, MA, 02139, USA

E-mail: naruminw@mit.edu

N. Nagaya, W. A. Tisdale

Department of Chemical Engineering, Massachusetts Institute of Technology, 77 Massachusetts Avenue, Cambridge, MA, 02139, USA

A. Alexiu, C. F. Perkinson, O. M. Nix, M. G. Bawendi, T. Van Voorhis

Department of Chemistry, Massachusetts Institute of Technology, 77 Massachusetts Avenue, Cambridge, MA, 02139, USA

**Abstract**

Singlet exciton fission has the potential to increase the efficiency of crystalline silicon solar cells beyond the conventional single junction limit. Perhaps the largest obstacle to achieving this enhancement is uncertainty about energy coupling mechanisms at the interfaces between silicon and exciton fission materials such as tetracene. Here, the previously reported silicon-hafnium oxynitride-tetracene structure is studied and a combination of magnetic-field-dependent silicon photoluminescence measurements and density functional theory calculations is used to probe the influence of the interlayer composition on the triplet transfer process across the hafnium oxynitride interlayer. It is found that hafnium oxide interlayers do not show triplet exciton sensitization of silicon, and that nitrogen content in hafnium oxynitride layers is correlated with enhanced sensitization. Calculation results reveal that defects in hafnium oxynitride interlayers with higher nitrogen content introduce states close to the band-edge of silicon, which can mediate the triplet exciton transfer process. Some defects introduce additional deleterious mid-gap states, which may explain observed silicon photoluminescence quenching. These results show that band-edge states can mediate the triplet exciton transfer process, potentially through a sequential charge transfer mechanism.

**1. Introduction**

The adoption of crystalline silicon (*c*-Si) photovoltaics is limited by the price of solar cells and the cost of their installation. Improving cell efficiency is an important goal because maximizing energy generation reduces the effective cost of both cells and installation. Singlet exciton fission (SF) has been proposed as a method for enhancing silicon solar cell efficiencies beyond the conventional theoretical limit for single junction devices.[1] It generates two triplet excitons from one singlet exciton.[2] If a SF material such as tetracene (Tc) is used to absorb the high energy photons of the solar spectrum, then transfer of the resulting triplet excitons to *c*-Si could increase silicon cell efficiencies from 29% to 35%–42%.[3–5]

Unfortunately, transfer of triplet excitons directly from Tc to *c*-Si has proven to be exceptionally challenging.[6–10] The fundamental obstacle is that Tc triplets are non-emissive states and incapable of near-field or radiative coupling to Si. Instead, triplet diffusion in molecular films typically relies on Dexter transport and involves simultaneous tunneling of the electron and hole from donor to acceptor molecules. Tunneling is inherently short range, limiting the thickness of silicon passivation layers, and increasing the impact of silicon surface defect states that quench triplet excitons.[11]

Previous bichromatic magnetic field-dependent measurements show that using a thin layer of hafnium oxynitride ($HfO_xN_y$) between Tc and n-type *c*-Si (n-Si) enables triplet exciton sensitization of silicon.[11] The sensitization effect is strongly dependent on the thickness of the $HfO_xN_y$ interlayer, with an optimum thickness of 8 Å, attributed to the interplay between carrier tunneling distance and silicon surface passivation. Both passivation and energy transfer processes are schematically summarized in **Figure 1**. $HfO_xN_y$ is expected to provide chemical passivation of dangling bonds at the silicon surface. The Si-$HfO_xN_y$-Tc samples also exhibit electric field-effect passivation when optically exciting both tetracene and/or silicon.[11] Trapping of minority (hole) carriers in the $HfO_xN_y$ interlayer is thought to be largely responsible for the electric field passivation effect. The trapped positive charge repels minority carriers from the surface of n-Si, reducing the surface recombination velocity. In contrast, electron traps only slightly affect the concentration of the majority carriers and have little effect on the rate of surface recombination.

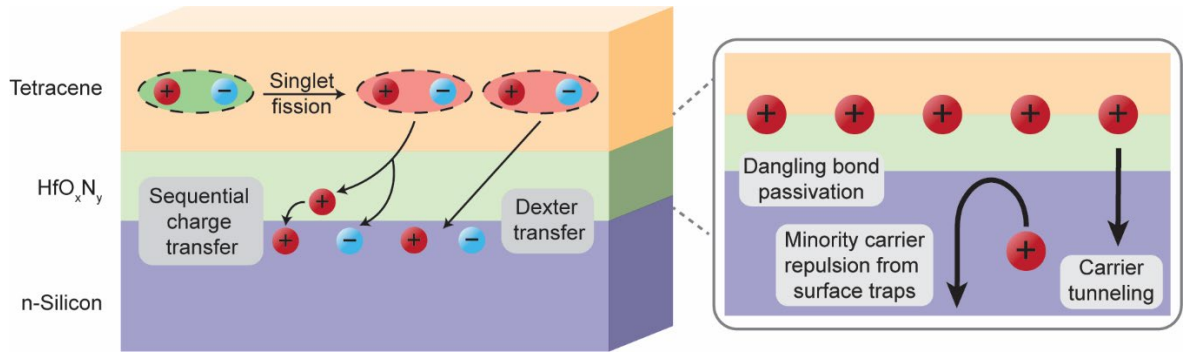

**Figure 1.** Schematic of the silicon-$HfO_xN_y$-tetracene structure studied in this work. Triplet excitons are formed from the singlet excitons in the tetracene layer through singlet fission. The triplet excitons can then either undergo a sequential charge transfer process or a Dexter transfer process to transfer to the *c*-silicon layer. The $HfO_xN_y$ interlayer provides chemical passivation by reacting with dangling bonds on the silicon surface. It also provides electric field-effect passivation of the silicon surface such that minority carriers in n-silicon are repelled from surface trap states. This passivation is proposed to be mediated by defect states.[11] The optimum interlayer thickness

is determined by the interplay between tunneling distance of the transferred carriers and silicon surface passivation.

Two potential mechanisms have been proposed to explain triplet exciton transfer from Tc to *c*-Si through a $HfO_xN_y$ layer.[11] Depicted in Figure 1, the triplet exciton could transfer through a sequential charge transfer mechanism, where the triplet exciton dissociates and the electron and hole transfer successively one after the other,[5] or through a Dexter transfer mechanism,[12] where the electron and hole transfer simultaneously to the silicon. Previous studies of LiF interlayers,[6,7] pyrene passivation layers[8] and covalently bound tetracene-derivative seed layers,[10] have not provided strong support for the effectiveness of Dexter transport at interfaces between molecules and *c*-Si.

The measured band alignment with the $HfO_xN_y$ interlayer[11] also does not appear to support the sequential charge transfer mechanism of the triplet excitons. The previous observation of electric field-effect passivation[11], however, points to the presence of defects in the $HfO_xN_y$ interlayer. In this work, we explore the potential role these defect states could play in the triplet exciton sensitization process by varying the composition of $HfO_xN_y$ interlayers and fabricating optical Si-$HfO_xN_y$-Tc samples to measure the interlayer-thickness-dependent silicon photoluminescence change of these samples under an external magnetic field. We correlate the experimental observations to density functional theory (DFT) calculations of defect state positions in these interlayers. Our results suggest that defect states in $HfO_xN_y$ are mediating sequential charge transfer of the triplet exciton.

**2. Results and Discussion**

$HfO_xN_y$ is a complex, non-stoichiometric material. To explore its properties and function at the interface between silicon and Tc, we compare different compositions of $HfO_xN_y$. The limiting case of hafnium oxide ($HfO_x$) is particularly important because $HfO_x$ has been shown to be a good passivating interlayer in Si solar cells, due to both chemical passivation of Si dangling bonds and field-effect passivation.[13–15] $HfO_x$ thin films can either be positively or negatively doped, depending on their stoichiometry.[15] In materials with a large concentration of oxygen vacancies, hafnia films develop a build-up of positive charge which electrically passivate n-doped Si.[13]

## 2.1. HfO$_x$N$_y$ Composition Variation

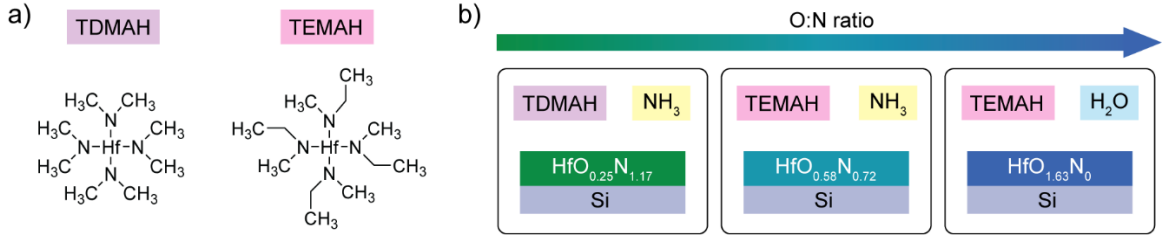

**Figure 2.** a) Hafnium precursors used to grow the HfO$_x$N$_y$ interlayers in this work: tetrakis(dimethylamino)hafnium (TDMAH) and tetrakis(ethylmethylamino)hafnium (TEMAH). b) Combinations of precursors used in the atomic layer deposition process for each layer and the corresponding compositions of the HfO$_x$N$_y$ films that were grown. The compositions were obtained using X-ray photoelectron spectroscopy on thicker films after etching *in-situ* to remove surface contaminants.

We grow HfO$_x$N$_y$ films on cleaned silicon surfaces using atomic layer deposition (ALD) with different precursors to achieve varying ratios of O:N. **Figure 2**a shows the chemical structures of the two hafnium precursors used in this work: tetrakis(dimethylamino)hafnium (TDMAH) and tetrakis(ethylmethylamino)hafnium (TEMAH). The HfO$_x$N$_y$ interlayers are grown following recipes for growth of Hf$_3$N$_4$.[11,16] The oxygen present in the films could come from residual oxygen in the reaction chamber, as well as from oxygen diffusion through post-deposition exposure of the films to air. TEMAH is reported to show unfavorable reactivity with the nitrogen precursor ammonia (NH$_3$).[16] Thus, we expect the HfO$_x$N$_y$ films grown using TEMAH to have a lower nitrogen content. We also prepare a HfO$_x$ film for comparison with no nitrogen content. The precursor combinations and compositions of the corresponding HfO$_x$N$_y$ films are presented in Figure 2b.

## 2.2. Optical Sample Characterization

To investigate the triplet sensitization efficiencies, we prepare optical samples of silicon with the deposited HfO$_x$N$_y$ interlayers of different thicknesses, and we then deposit 30 nm of Tc on top. We characterize the optical samples by measuring the silicon photoluminescence (PL) change under an applied magnetic field.

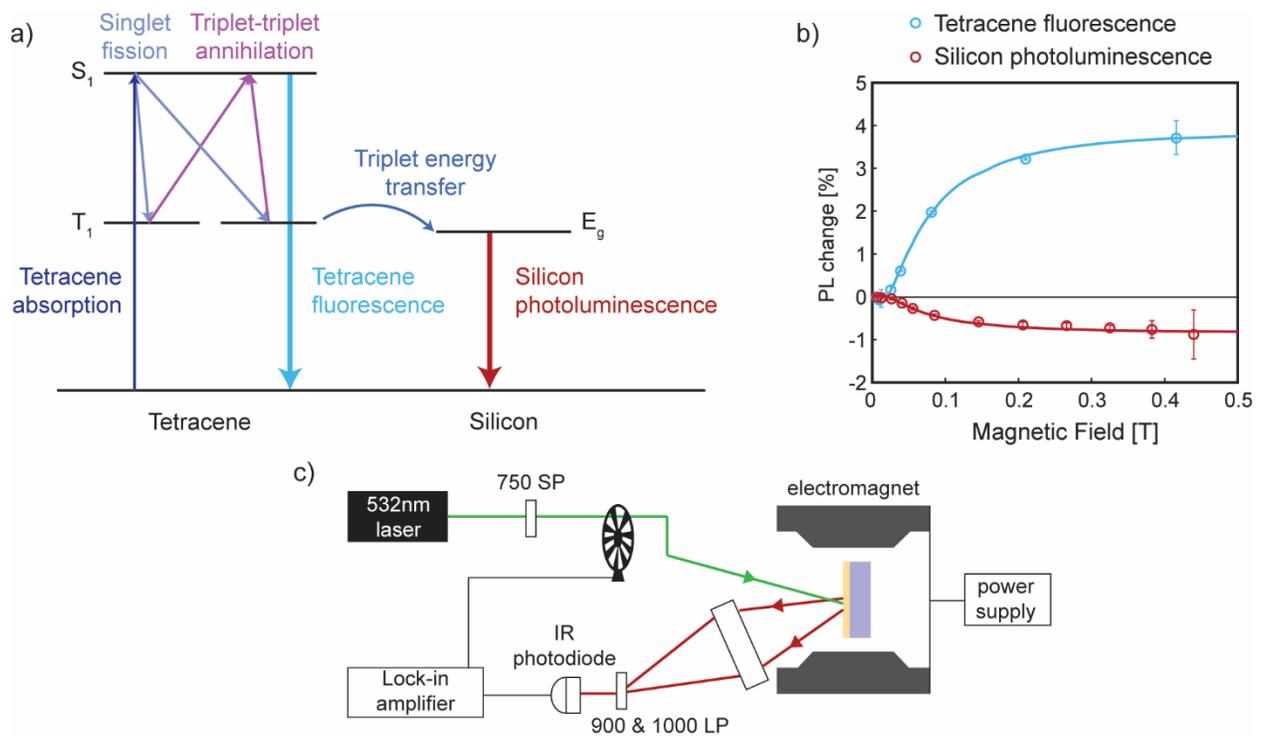

**Figure 3**. a) Schematic showing the processes that can occur in the silicon-$HfO_xN_y$-tetracene optical sample. Tetracene can absorb photons to generate singlet excitons that can undergo singlet fission to form two triplet excitons each. The triplet excitons could either combine via triplet-triplet annihilation to form a singlet exciton, or transfer to the silicon. Tetracene singlet excitons can decay to the ground state to emit fluorescence. The excited silicon can decay to the ground state to emit photoluminescence. b) Photoluminescence change measured as a function of magnetic field for tetracene fluorescence and for silicon photoluminescence. The solid lines are fits to the singlet fission characteristic curve in tetracene. c) The experimental setup used to measure the silicon photoluminescence change as a function of magnetic field.

When the Tc layer is excited, it absorbs photons to generate singlet excitons. The singlet exciton can undergo singlet fission to generate a correlated triplet pair state with spin-singlet character, which either diffuses into separated triplet excitons or recombines through triplet-triplet annihilation to form a singlet exciton. Triplet excitons in Tc can also transfer their energy to silicon. **Figure 3**a shows a summary of the processes that can occur in the silicon-$HfO_xN_y$-tetracene optical samples. The coupling of the singlet exciton and the correlated triplet pair state can be modulated by a magnetic field, affecting the equilibrium population of singlet and triplet excitons, as described by the Merrifield model.[17] At low fields, the overall singlet fission rate is increased, followed by a decrease at high fields. This results in a characteristic response of the steady-state Tc fluorescence signal to an external magnetic field, as shown in Figure 3b.[18] Measuring the silicon PL shows the same characteristic response but inverted, which aligns with the expectation that triplet excitons from Tc are sensitizing the Si. It should be noted that the sensitization effects potentially include both energy transfer and electric field-effect passivation induced by defect charging at the interface.[11]

To infer the efficiency of triplet exciton sensitization of Si, we measure the percentage change in silicon PL from the Si-HfO$_x$N$_y$-Tc samples upon application of a 0.4 T magnetic field. The samples are excited by a 532 nm laser source and the silicon PL is captured and focused onto an IR photodetector. The experimental setup is depicted in Figure 3c.

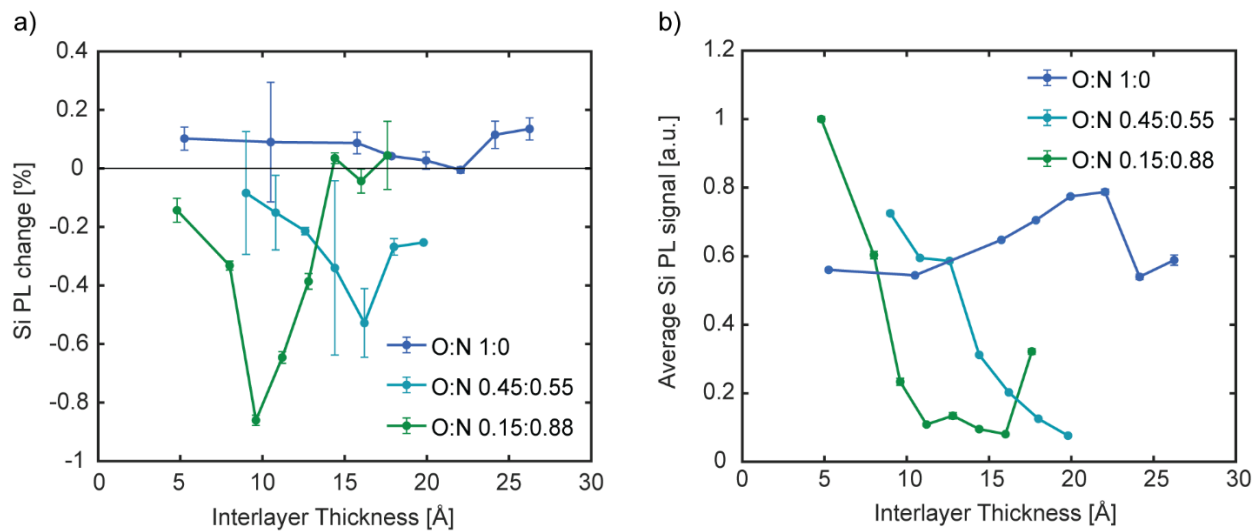

**Figure 4.** a) Percentage silicon photoluminescence (PL) change of Si-HfO$_x$N$_y$-Tc samples upon application of a 0.4 T magnetic field at different HfO$_x$N$_y$ interlayer thicknesses and compositions. The magnitude of the silicon PL change increases with higher nitrogen content in the HfO$_x$N$_y$ layer. b) Average silicon PL signal of Si-HfO$_x$N$_y$-Tc samples at different HfO$_x$N$_y$ interlayer thicknesses and compositions. The average silicon PL is roughly constant with thickness for HfO$_x$, but trends downwards with increasing thickness of the HfO$_x$N$_y$ interlayers.

Previous work on HfO$_x$N$_y$ layers found a strong thickness-dependence of the sensitization effect.[11] As a result, we measure the interlayer thickness-dependent silicon PL change for the samples with different interlayer compositions. The results are presented in **Figure 4**a). Notably, changing the interlayer composition affects both the magnitude of the silicon PL change and the thickness-dependence. The $HfO_x$ films with no nitrogen showed no negative Si PL change with applied field at all thicknesses, implying no triplet exciton sensitization of the silicon. With increasing nitrogen content in the $HfO_xN_y$ interlayer, the samples exhibit a higher magnitude of silicon PL change with an applied magnetic field, implying a greater sensitization effect. The optimum thickness also appears to decrease with greater nitrogen content.

We also measure the average silicon PL signal as a function of interlayer thickness and composition (Figure 4b). We again observe differences in the trends between samples with the $HfO_x$ and $HfO_xN_y$ interlayers. Unlike the thickness-dependence of the samples with the $HfO_x$ interlayers, the samples with both $HfO_xN_y$ interlayers show a nearly monotonically decreasing silicon PL signal with increasing interlayer thickness.

The quenching of silicon PL in the $HfO_xN_y$ interlayers is an indicator of mid- and near-band-edge defect states. In combination with observed enhanced triplet sensitization, we propose that triplet excitons are transferred through a sequential charge transfer mechanism, mediated by defect states in the $HfO_xN_y$ layer.

### 2.3. Computational investigation of $HfO_xN_y$

To evaluate the hypothesis that defect states near the band edge of silicon may support sequential charge transfer of the triplet excitons, we perform density functional theory (DFT) calculations to assess the density of states (DOS) for a range of materials of interest.

We start with hafnium oxide ($HfO_2$), a well-studied wide band-gap metal oxide. Given that polycrystalline phases only appear after annealing at high temperatures (> 500 °C),[19,20] most hafnia thin films reported in the literature exhibit amorphous structures. A lack of crystallinity precludes us from making direct comparisons with experiments. Therefore, we choose to focus on the thermally accessible phases of $HfO_2$, which are the monoclinic ground state (*P2₁/c*) and the orthorhombic state (*Pbca*), which is 11.1 meV/atom higher in energy (see Figure S5 for energy diagram). A brief analysis of point defects in monoclinic $HfO_2$ is reported in Figure S6, which agrees with previous computational and experimental results.[21–23] Overwhelmingly, these studies show that O vacancies in $HfO_2$ can act as electron traps, and defects associated with N atoms may cause additional trap states. Note that our calculations employ the global hybrid PBE0,[24,25] which dramatically improves semiconductor band gap and level alignment compared to the generalized gradient approximation (GGA) used in earlier work. However, at least in the case of $HfO_2$, the qualitative conclusions ae unchanged.

To extend the investigation to $HfO_xN_y$, we carry out a systematic interpolation between $HfO_2$ and $Hf_3N_4$, to establish the likelihood that different O:N ratios will be conducive to efficient triplet exciton transfer from Tc to Si. For a smooth interpolation, the two endmembers should have similar crystal structures, which leads us to focus on orthorhombic $HfO_2$ (*Pbca*) and $Hf_3N_4$ (*Pnma*), as there is no thermally accessible monoclinic $Hf_3N_4$.

Oxygen atoms were progressively substituted by nitrogen atoms, maintaining the overall stoichiometry, to yield a range of compositions of the form $HfO_xN_y$, with N substitution percentages varying from 13-91% (see **Figure 5**a for a representative unit cell). The Si band gap and offset between the $HfO_xN_y$ and Si valence band edges are assumed to be 1.1 eV and 2.4 eV, respectively, as calculated from ultraviolet photoelectron spectroscopy experiments by Einzinger *et al*.[11] Other experimental works obtain a range of values between 2.4-3 eV[26–30] for the valence band offset, depending on the exact deposition setup and any post-deposition annealing treatments. We highlight that our computational results will be somewhat dependent on this exact value, however we choose 2.4 eV as the valence band offset for consistency with the experiments of Einzinger *et al*.[11]

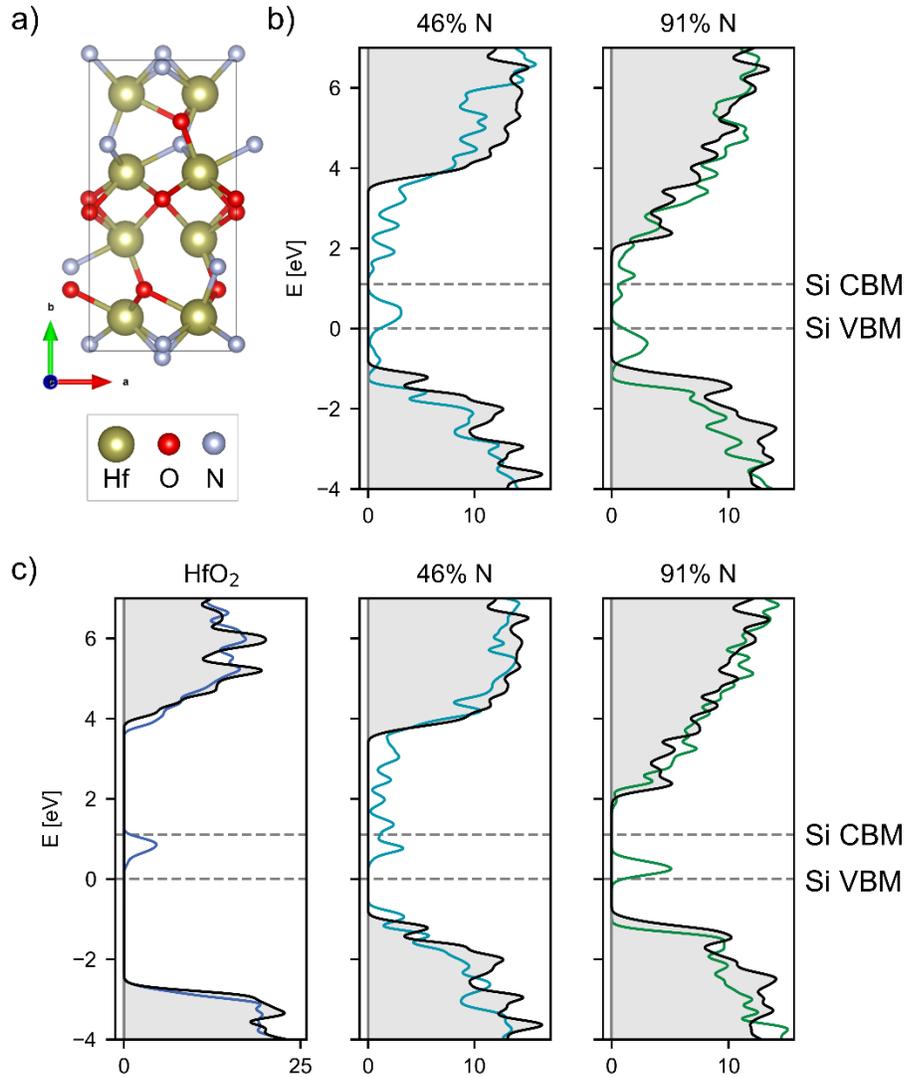

**Figure 5.** a) Representative unit cell of $HfO_xN_y$ (with 46% N). b) Density of states plots for N vacancy defects in two representative $HfO_xN_y$ compositions. The DOS of the pristine unit cell (grey shaded region) is compared with the defective DOS (colored). The Si valence band maximum (VBM) and conduction band minimum (CBM) are marked by dotted grey lines. The energy axis is shifted such that the Si VBM is at 0 energy. c) DOS plots for O vacancy defects in $HfO_2$ and $HfO_xN_y$.

In agreement with previous experimental and computational results,[31–33] the band gap of hafnium oxynitride decreases monotonically with increasing N content (see Figure S4). The band gaps of all pristine samples are free of any energy states that could realistically mediate sequential charge transfer of triplet excitons to silicon.

### 2.3.1. Defective structures

We introduced a variety of point defects: O, N, and Hf vacancies. All defects considered are charge neutral, to remove the dependence of their formation energy on the Fermi level. An analysis of charged defects could make the object of a further study. Hf vacancies form no mid-gap states of interest, so we reserve their discussion for the Supplementary Information (Figure S9).

Representative DOS plots are shown in Figure 5b for N vacancies in two $HfO_xN_y$ compositions, a more N poor case (46% N) that exhibits a trap state in the middle of the Si band gap and a N rich case (91% N) where the defect state could conceivably trap holes. Similarly for O vacancies, a comparison is shown between the same two $HfO_xN_y$ compositions and $HfO_2$ in Figure 5c.

One possible pathway for sequential charge transfer mediated by a defect state in $HfO_xN_y$ is presented in **Figure 6**a, where the electron is initially transferred to the silicon conduction band ($E_C$) and the hole is supported at a $HfO_xN_y$ defect state ($E_D$) near the valence band edge of silicon. The hole is then subsequently transferred to the silicon valence band ($E_V$).

Such a defect state needs to have significant density of states close to the Si valence band edge, and ideally, the energy of the charge-separated state ($E_\pm$) should be lower than the triplet energy in Tc (~1.25 eV, within thermal energy)[11]: $E_\pm \leq E_{T,Tc}$. We assume binding energy can be neglected for the charge-separated state energy due to expected charge screening in silicon.[34,35] Furthermore, the energy barrier for the subsequent hole transfer from $HfO_xN_y$ to Si must be sufficiently low, on the order of thermal energy at room temperature, to avoid permanent trapping of the minority carriers in defect states.

A summary of all vacancy energy levels is shown in Figure 6b, for $HfO_2$, $HfO_xN_y$, and $Hf_3N_4$. Only the maximum of each defect peak in the DOS plots is shown, although each peak is in reality broadened by up to 0.5 eV each (see Figure 5b and c for examples of realistic DOS plots, as well as Figures S7 – S11 for DOS plots of all possible defects).

O vacancies cause electron traps for low N compositions, shifting to potential hole traps as the N content increases. N vacancies generally lead to the formation of hole traps; for higher N content, the defect states can even lie slightly below the Si VBM, which could promote a barrierless hole transfer from the defect energy level to the silicon VBM (see Figure 6a). The overall trend is for vacancy defect levels to decrease in energy as the N content increases, with the optimal composition for hole trap formation being higher than 67% N.

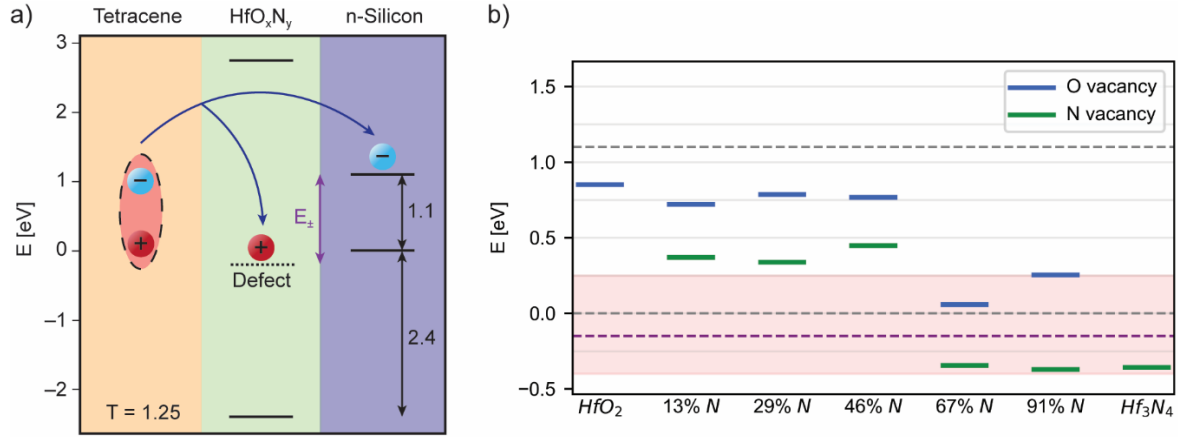

**Figure 6.** a) Schematic showing the energy level alignment of the silicon-HfO$_x$N$_y$-tetracene structure, as previously reported.[11] Overlaid, one possible example of a defect state in HfO$_x$N$_y$ helping mediate the triplet exciton dissociation and sequential charge transfer process of the triplet exciton from tetracene. This forms an intermediate charge-separated state with energy E$_\pm$ (assuming negligible binding energy[34,35]), where the electron is on the silicon conduction band edge and the hole is on the defect state level in HfO$_x$N$_y$. b) Summary of defect energy levels caused by O vacancies (blue) and N vacancies (green). The Si valence band maximum (VBM) and conduction band minimum (CBM) are marked with dotted grey lines, and the dotted purple line represents the minimum energy a defect state could have such that $E_\pm \leq E_{T,Tc}$. The red shaded box marks the energy region for potential hole traps. The energy axis is shifted such that the Si VBM is at 0 energy.

### 2.3.2. Defect formation energies

The relative alignment of potential trap states offers an explanation for which types of defects could be responsible for the success of using the HfO$_x$N$_y$ interlayer in singlet fission solar cells, but gives no information about the prevalence of such defects. Computing their formation energies provides some insight into their relative thermodynamic stabilities and the likelihood of defect formation in a real material.

Defect formation energies and concentrations are highly dependent on the conditions under which these materials are synthesized, most notably the experimental temperature and the chemical potentials, $\mu_N$ and $\mu_O$, of N and O. Making reasonable approximations to $\mu_N$ and $\mu_O$ (see computational details) leads to the formation energies shown in **Figure 7**. We see that for all O:N ratios, oxygen vacancies are more stable than nitrogen vacancies, which would in turn indicate that we should expect oxygen vacancies to be more common. As implied by the low or even negative formation energies, oxygen vacancies form readily at any composition (negative formation energies suggest spontaneous defect formation). The second observation is that nitrogen vacancies are systematically more stable for higher nitrogen composition as compared to lower, suggesting that more nitrogen-rich $HfO_xN_y$ should have more nitrogen defects.

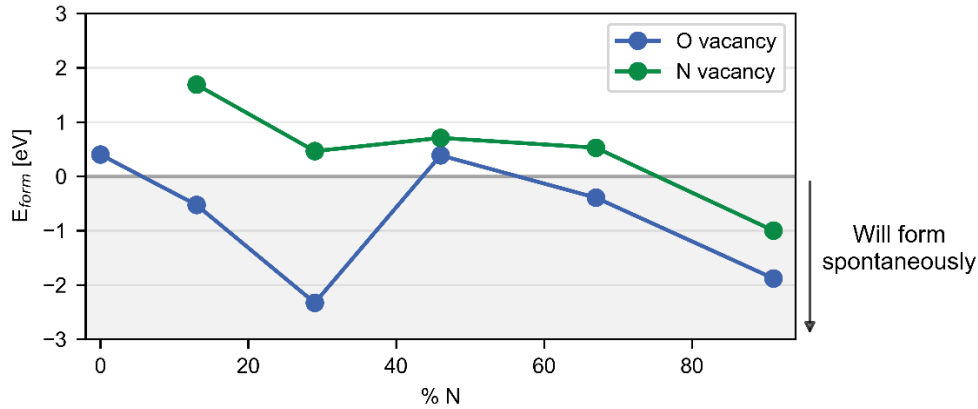

**Figure 7.** Formation energies for O vacancies (blue) and N vacancies (green), for $HfO_2$ and $HfO_xN_y$ compositions. O vacancies form more readily than N vacancies, with both forming spontaneously at the 91% N composition.

Using a temperature of 423 K, in line with the reaction chamber temperature used in our experimental work, and assuming the structure is thermodynamically (rather than kinetically) determined, we can also estimate defect concentrations using an Arrhenius-like expression (Table T5 in Supplementary Information). Oxygen vacancy concentrations are large, irrespective of the $HfO_xN_y$ composition, estimated to be on the order of $10^{16}$ cm$^{-3}$ in $HfO_2$. This estimate is in accordance with previous computational[36] and experimental[37,38] studies, and may even be underestimating the concentration of oxygen vacancies in a real material. The highest nitrogen vacancy concentrations are obtained for the material with 91% N ($10^{31}$ cm$^{-3}$). These quantitative predictions confirm the qualitative expectations implied by the defect-induced energy levels in Figure 6b.

This computational analysis suggests that $HfO_xN_y$ with higher N content supports the formation of defect states near the valence band edge of silicon, helping to mediate the sequential charge transfer mechanism proposed in Figure 6a. This hypothesis is consistent with the experimentally observed lower optimum transfer thickness and higher silicon PL change magnitude for $HfO_xN_y$ films with higher N concentration. Additionally, the defect formation energy calculations suggest that $HfO_xN_y$ films with higher N concentrations could favor the formation of N vacancies compared to the films with moderate N concentrations.

For the $HfO_x$ layer, calculations suggest that oxygen vacancies could result in defect states near the conduction band of silicon, which could support triplet exciton transfer through an initial hole transfer. However, we note that the magnetic field-dependent measurements detect both energy transfer and passivation. The wafers used in the study are n-doped with holes as the minority carriers. Thus, the electrons in the $HfO_x$ layer create an electric field that attracts the minority carriers to the surface, resulting in recombination of any transferred triplet excitons.

Our calculations also support the observation of nearly monotonically decreasing silicon PL signal with increasing $HfO_xN_y$ interlayer thickness, consistent with the increased number of mid-gap defect-induced states in thicker interlayers from nitrogen and oxygen. The defect formation studies for the $HfO_x$ layer also suggest the presence of fewer mid-gap defect states, consistent with the experimentally-observed relatively constant silicon PL signal with thickness.

## 3. Conclusion

In conclusion, we show that sequential charge transfer of triplet excitons from tetracene to crystalline silicon can be mediated through defect states in the $HfO_xN_y$ interlayer. Our results suggest that $HfO_xN_y$ films with higher nitrogen content are more likely to form N vacancy defects, introducing a density of near-band-edge states that lie below the silicon valence band maximum. These states correlate with increased magnitudes of magnetic field-induced modulation of silicon photoluminescence and reduced optimum interlayer thickness for triplet sensitization of silicon. We also find, however, that defects in $HfO_xN_y$ interlayers introduce many deep mid-gap states which can quench the overall silicon photoluminescence. Future efforts for achieving high efficiency singlet fission-sensitized silicon photovoltaics will require interlayers that account for the importance of both near-band-edge states and electric field passivation of the silicon surface.

## 4. Methods

### 4.1 Experimental section

*Materials*

n-doped prime-grade single-side-polished silicon wafers with 0.25-0.75 Ωcm resistivity were purchased from UniversityWafer. Tetracene was purchased from Sigma Aldrich (99.99% purity) and purified via sublimation in a tube furnace once.

*Sample Preparation*

The silicon wafers were diced into 1-inch squares and cleaned by sonicating in detergent solution (Micro-90), deionized water, and acetone, followed by immersion in boiling isopropanol. The wafers were then dried with pressurized nitrogen and transferred to a clean room. We performed a standard Radio Corporation of America (RCA) cleaning process on the wafers, which left a thin oxide layer on the silicon surface. Following the clean, the wafers were quickly transferred for atomic layer deposition of the $HfO_xN_y$ layers (Cambridge Nanotech Savannah). The base pressure of the chamber was 0.4 torr at a nitrogen flow of 40 $cm^3$ $min^{-1}$.

For the $HfO_{0.25}N_{1.17}$ layer, tetrakis(dimethylamino)hafnium (TDMAH) was used as the Hf precursor and ammonia was used as the nitrogen precursor, with oxygen likely from residual oxygen in the chamber and from film exposure to air post-deposition. The TDMAH precursor was heated to 75 °C and the ALD reaction chamber was set to 150 °C. The TDMAH precursor was pulsed for 30 ms, and the $NH_3$ precursor was pulsed for 15 ms for every cycle.

For the $HfO_{0.58}N_{0.72}$ layer, tetrakis(ethylmethylamino)hafnium (TEMAH) was used as the Hf precursor and $NH_3$ was used as the nitrogen precursor, with oxygen likely from residual oxygen in the chamber and from film exposure to air post-deposition. The TEMAH precursor was heated to 110 °C and the ALD reaction chamber was set to 150 °C. The TEMAH precursor was pulsed for 400 ms, and the $NH_3$ precursor was pulsed for 15 ms for every cycle.

For the $HfO_{1.63}$ layer, TEMAH was used as the Hf precursor and water was used as the oxygen precursor. The TEMAH precursor was heated to 110 °C and the ALD reaction chamber was set to

200 °C. The TEMAH precursor was pulsed for 250 ms, and the H$_2$O precursor was pulsed for 15 ms for every cycle.

After deposition of the HfO$_x$N$_y$ interlayer, the samples were quickly transferred to a vacuum chamber at a pressure of $< 1\times10^{-6}$ torr for tetracene deposition via thermal evaporation. 30 nm of tetracene was evaporated at a rate of 1 Å/s. The samples were then encapsulated with a glass slide and ultraviolet curable epoxy in a dry nitrogen atmosphere ($< 1$ ppm O$_2$). A square foil piece was used to cover the active area during UV exposure.

*X-ray Photoelectron Spectroscopy (XPS)*

XPS measurements were performed using a PHI VersaProbe II X-ray photoelectron spectrometer with monochromated Al K-α X-rays. Compositional depth-profiles were obtained by sputtering the surface with C$_{60}$ ions (operated at 2 kV, 1 µA, over a 2 x 2-mm area) and measuring the photoelectron peak areas for C1s, N1s, O1s, Si2p, Hf4f in 1-minute intervals. The reported compositions were obtained after 5 minutes of sputtering on 25 nm-thick films to remove surface impurities and obtain compositions close to the sample surface.

*Magnetic-Field-Dependent Photoluminescence Measurements*

Optical samples were placed in Voigt geometry (perpendicular to the magnetic field) between the poles of an electromagnet and excited by a mechanically-chopped (281 Hz) 200 mW 532-nm laser (Coherent Verdi G18). The tetracene fluorescence was collected by a silicon photodetector (Newport 818-SL) connected to a lock-in amplifier (Stanford Research Systems SR 830), with a 532 nm notch filter and a 550 nm longpass filter in front of the detector. The silicon photoluminescence was collected by an IR photodetector (Newport 818-IR), with 900 nm and 1000 nm longpass filters in front of the detector to ensure collection of only the silicon photoluminescence. The electromagnet was periodically switched on and off at different field strengths for 4-cycle loops at each field strength to obtain the measurements in Figure 6b. For the thickness-dependent magnetic-field-dependent measurements, the above experiments were conducted for a magnetic field strength of 0.4 T, as measured by a gaussmeter (Lakeshore HMMT-6J04-VF).

**4.2. Computational Methods**

All calculations are performed using density functional theory at the hybrid functional level (PBE0), using the 'light' default settings for basis sets and numerical integration grid settings, as defined in the FHI-aims software.[39–42] Following convergence tests (see Figure S9), we chose a 2×2×2 k-point grid for the geometry optimizations, and a 4×4×4 grid for the density of states calculations. For a tightly converged geometry optimization, we used a threshold value of $5 \times 10^{-3}$ eV/Å for the magnitude of forces acting on nuclei. Scalar relativistic corrections were employed, using the atomic zeroth order regular approximation (ZORA).[39] 16- and 28-atom unit cells were used for $HfO_2$ and $Hf_3N_4$, respectively.

The procedure for computing defect formation energies and concentrations is well-established,[22,43,44] and we describe the main features here. The expression for the formation energy is:

$$E_{form} = E_{defect} - E_{pristine} + \sum_i n_i \mu_i \qquad (1)$$

where $n_i$ is the number of atoms added or removed to form the defect, and $\mu_i$ is the chemical potential. Since we are only considering neutral defects, there is no need to account for variations in the Fermi energy, or to apply electrostatic corrections to account for image charge interactions.

The chemical potentials are not constant, instead varying between certain limits defined by the thermodynamics of the system. O- and N-rich limits are defined by:

$$\mu_N \leq \frac{\mu_{N_2}}{2}, \mu_O \leq \frac{\mu_{O_2}}{2} \qquad (2)$$

The enthalpies of formation of stoichiometric $HfO_2$ and $Hf_3N_4$ place additional constraints on the chemical potentials, defining the O- and N-poor limits, respectively:

$$\Delta E_f^{HfO_2} = \mu_{HfO_2} - \mu_{Hf} - 2\mu_O, \Delta E_f^{HfO_2} \leq 0 \qquad (3)$$

$$\Delta E_f^{Hf_3N_4} = \mu_{Hf_3N_4} - 3\mu_{Hf} - 4\mu_N, \Delta E_f^{Hf_3N_4} \leq 0 \qquad (4)$$

For results reported in the main text, we use the O-poor limit, in line with described experimental conditions (only residual oxygen is present in the ALD chamber). Similar synthetic procedures for $HfO_2$ lead to high concentrations of oxygen vacancies[37] and sub-stoichiometric Hf:O ratios of 1.91 or less.[45] As for nitrogen, we choose an average chemical potential between the N-rich and

N-poor limits. A full summary of all possible limits is included in the Supplementary Information (Tables 1 - 4).

Estimates of defect concentrations can be obtained using an Arrhenius-like expression, assuming a deposition temperature of 423 K (similar to the experimental ALD temperature):

$$\boldsymbol{n = n_0 e^{-\frac{E_{form}}{kT}}} \tag{5}$$

Using the methods outlined previously, we obtained $\boldsymbol{\Delta E_f^{HfO_2}} = -11.937$ eV/unit cell, in line with the experimental enthalpy of formation of $-11.864$ eV/unit cell.[46]

**Supporting Information**

Supporting Information is available from the Wiley Online Library or from the author.

**Acknowledgements**

This work was primarily supported by the US Department of Energy, Office of Science, Office of Basic Energy Sciences, Materials Chemistry Program through award number DE-FG02-07ER46454 (N.N., A.A., C.F.P., O.M.N., T.V.V., M.A.B.). C.F.P. was partially supported by the National Science Foundation Graduate Research Fellowship under Grant No. 1122374. W.A.T. and N.N. were partially supported by the US Department of Energy, Office of Science, Office of Basic Energy Sciences, Physical Behavior of Materials Program through award number DE-SC0019345. A.A. was partially supported by the Haslam & Dewey Summer Fellowship.

Received: ((will be filled in by the editorial staff))
Revised: ((will be filled in by the editorial staff))
Published online: ((will be filled in by the editorial staff))

Supporting Information

**Triplet Exciton Sensitization of Silicon Mediated by Defect States in Hafnium Oxynitride**

*Narumi Nagaya\*, Alexandra Alexiu, Collin F. Perkinson, Oliver M. Nix, Moungi G. Bawendi, William A. Tisdale, Troy Van Voorhis, Marc A. Baldo*

**X-ray photoelectron spectroscopy of hafnium oxynitride and hafnium oxide films**

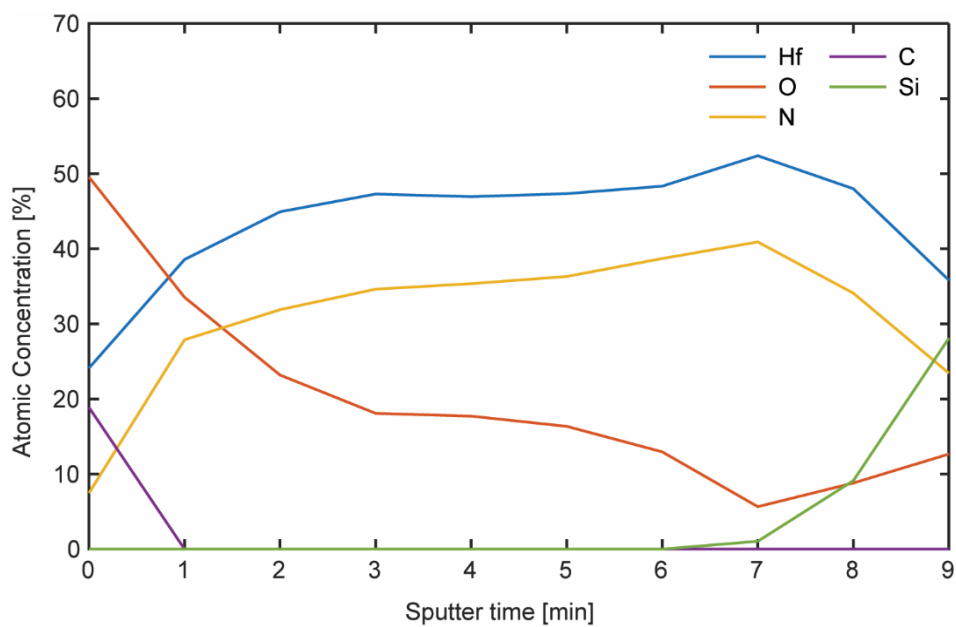

**Figure S1.** Compositional depth-profiles obtained from X-ray photoelectron spectroscopy measurements on the HfO$_x$N$_y$ film deposited using tetrakis(dimethylamino)hafnium (TDMAH) and NH$_3$ precursors. C$_{60}$ ions were used to sputter the surface, and the photoelectron peak areas for C1s, N1s, O1s, Si2p, Hf4f were measured in 1-minute sputtering intervals.

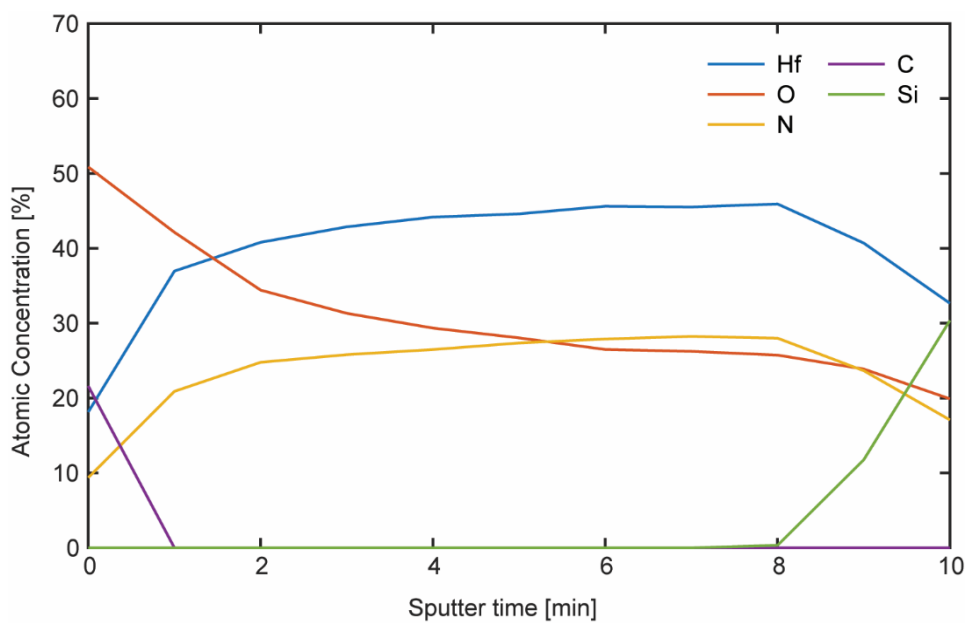

**Figure S2.** Compositional depth-profiles obtained from X-ray photoelectron spectroscopy measurements on the HfO$_x$N$_y$ film deposited using tetrakis(ethylmethylamino)hafnium (TEMAH) and NH$_3$ precursors. C$_{60}$ ions were used to sputter the surface, and the photoelectron peak areas for C1s, N1s, O1s, Si2p, Hf4f were measured in 1-minute sputtering intervals.

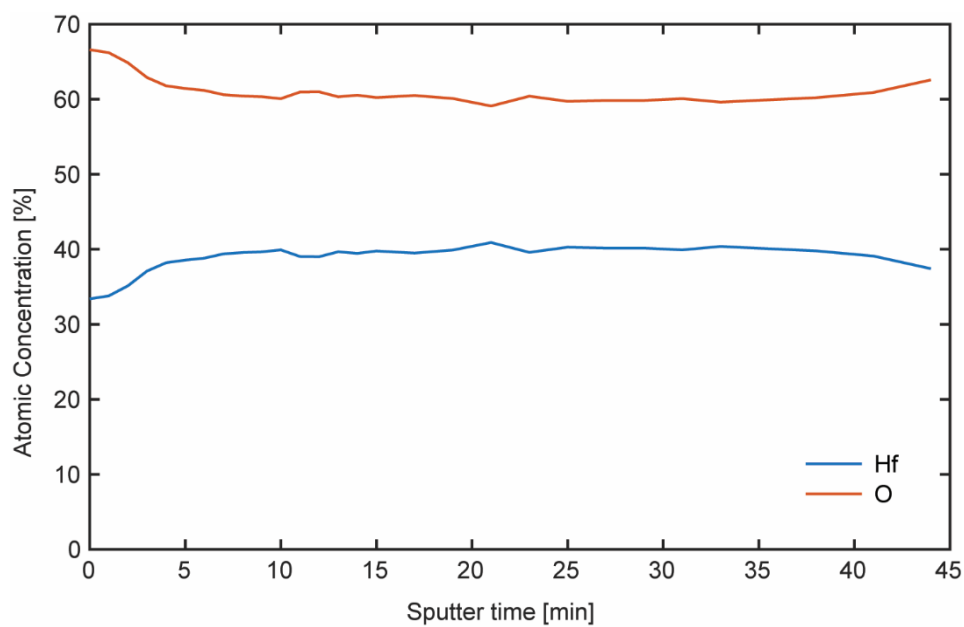

**Figure S3.** Compositional depth-profiles obtained from X-ray photoelectron spectroscopy measurements on the HfO$_x$ film deposited using tetrakis(ethylmethylamino)hafnium (TEMAH) and H$_2$O precursors. C$_{60}$ ions were used to sputter the surface, and the photoelectron peak areas for O1s, Hf4f were measured in 2-minute sputtering intervals.

**Additional computational investigations of hafnium oxynitride and hafnium oxide films**

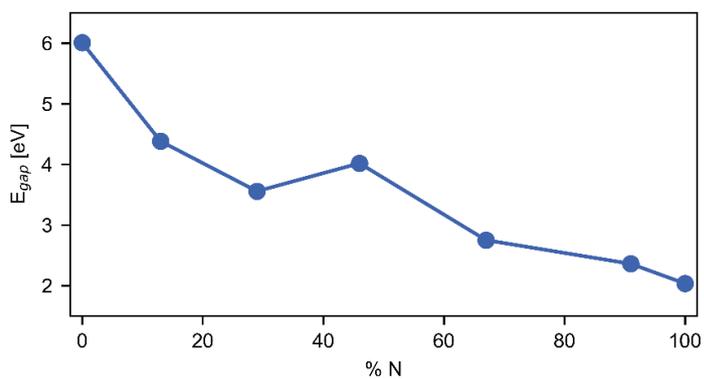

**Figure S4.** Band gap variation with N composition.

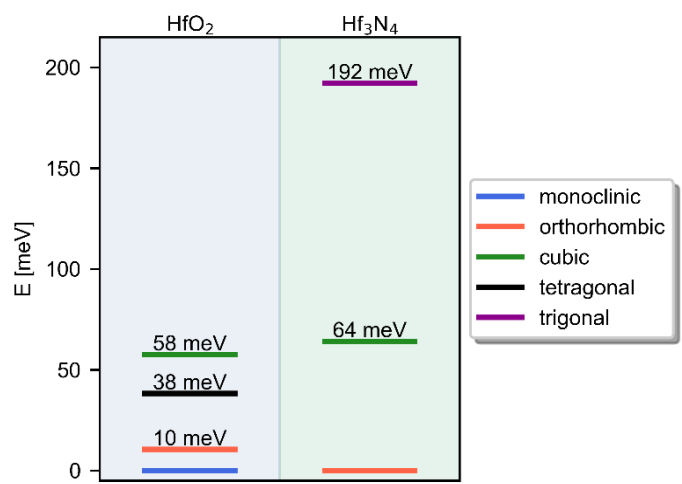

**Figure S5.** Relative energies of $HfO_2$ and $Hf_3N_4$ phases.

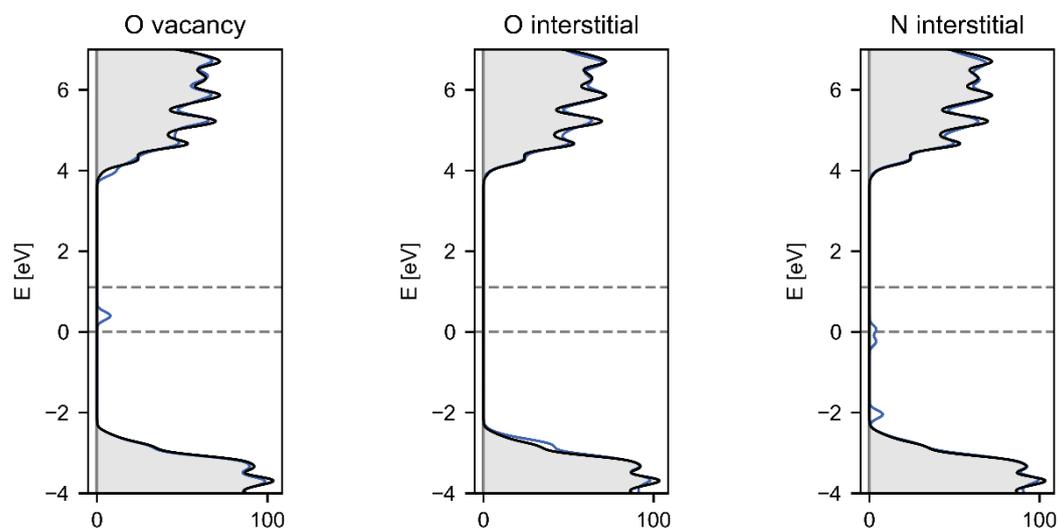

**Figure S6.** Density of states plots for O vacancy, O interstitial, and N interstitial defects in monoclinic $HfO_2$. The DOS of the pristine unit cell (grey shaded region) is compared with the defective DOS (colored). The Si valence band maximum (VBM) and conduction band minimum (CBM) are marked by dotted grey lines. The energy axis is shifted such that the Si VBM is at 0 energy.

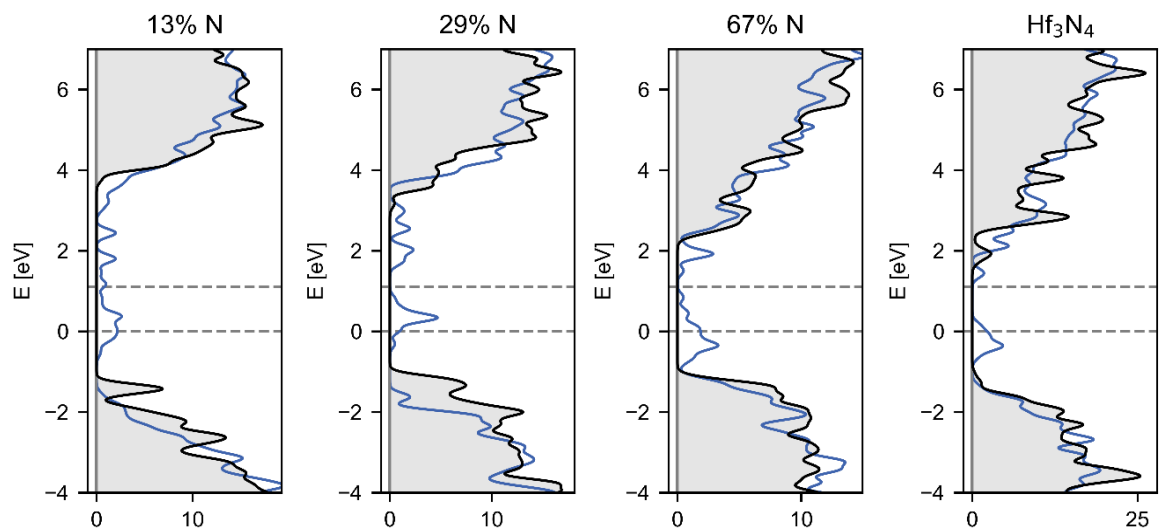

**Figure S7.** Density of states plots for N vacancy defects in a range of $HfO_xN_y$ compositions.

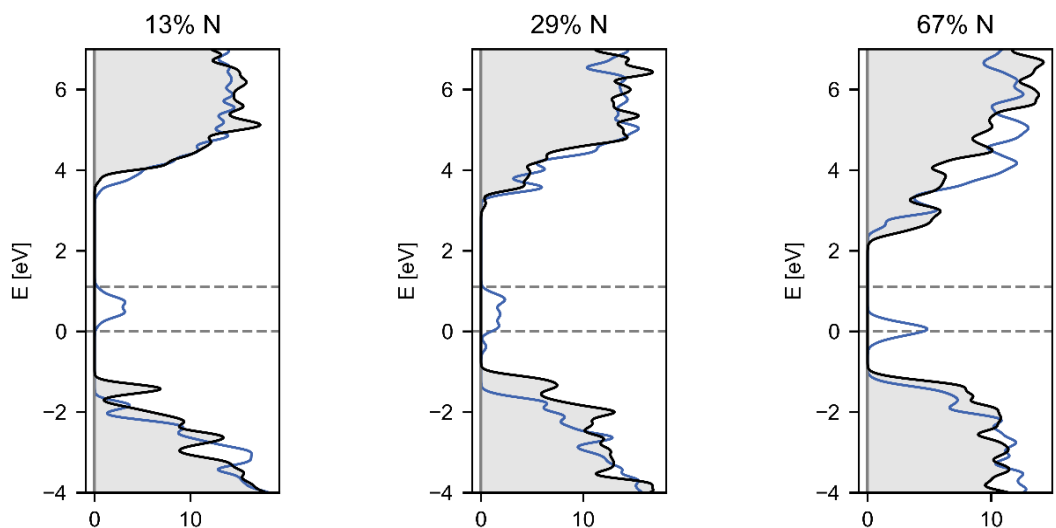

**Figure S8.** Density of states plots for O vacancy defects in a range of $HfO_xN_y$ compositions.

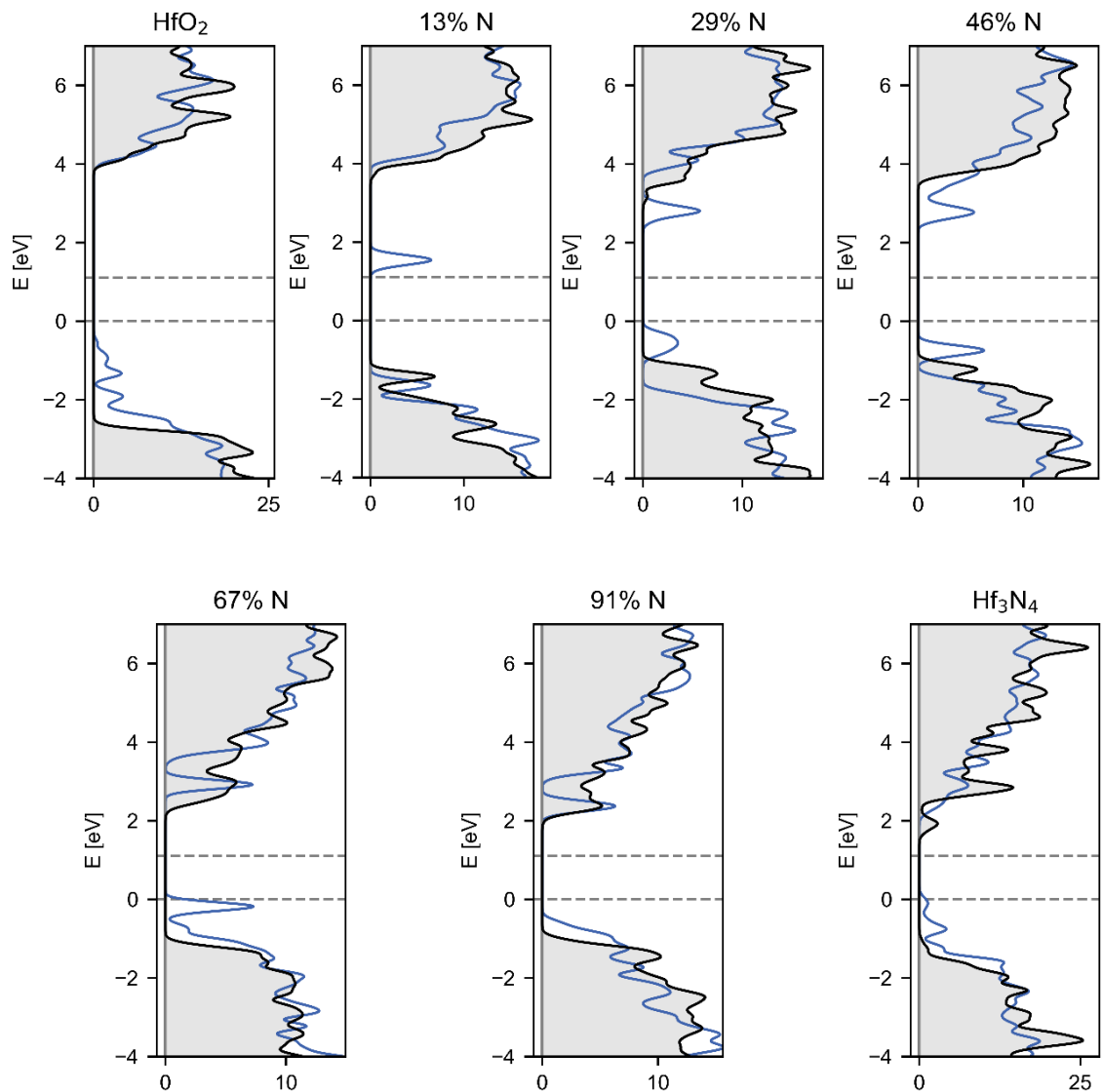

**Figure S9.** Density of states plots for Hf vacancy defects in a range of $HfO_xN_y$ compositions.

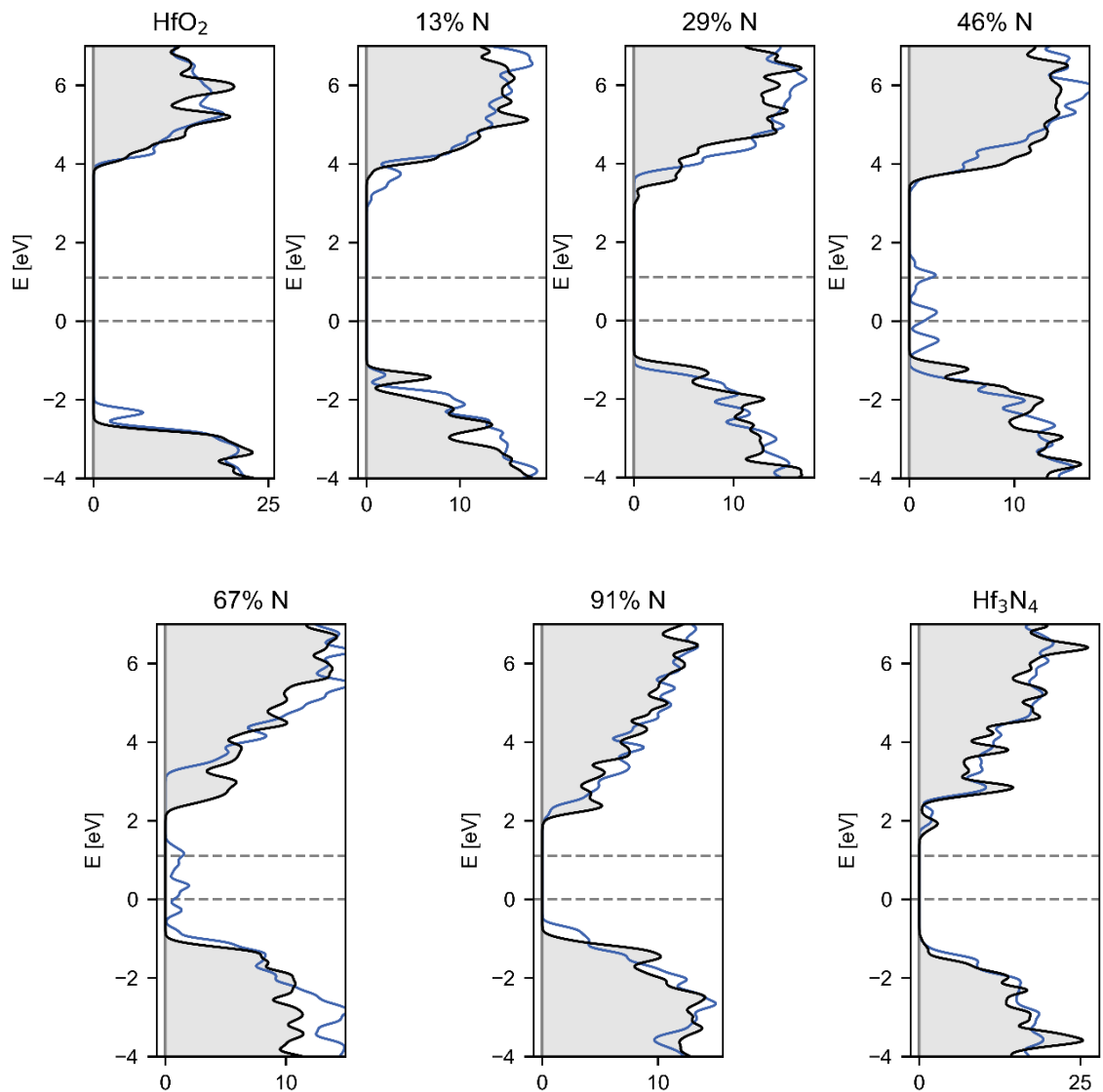

**Figure S10.** Density of states plots for O interstitial defects in a range of $HfO_xN_y$ compositions.

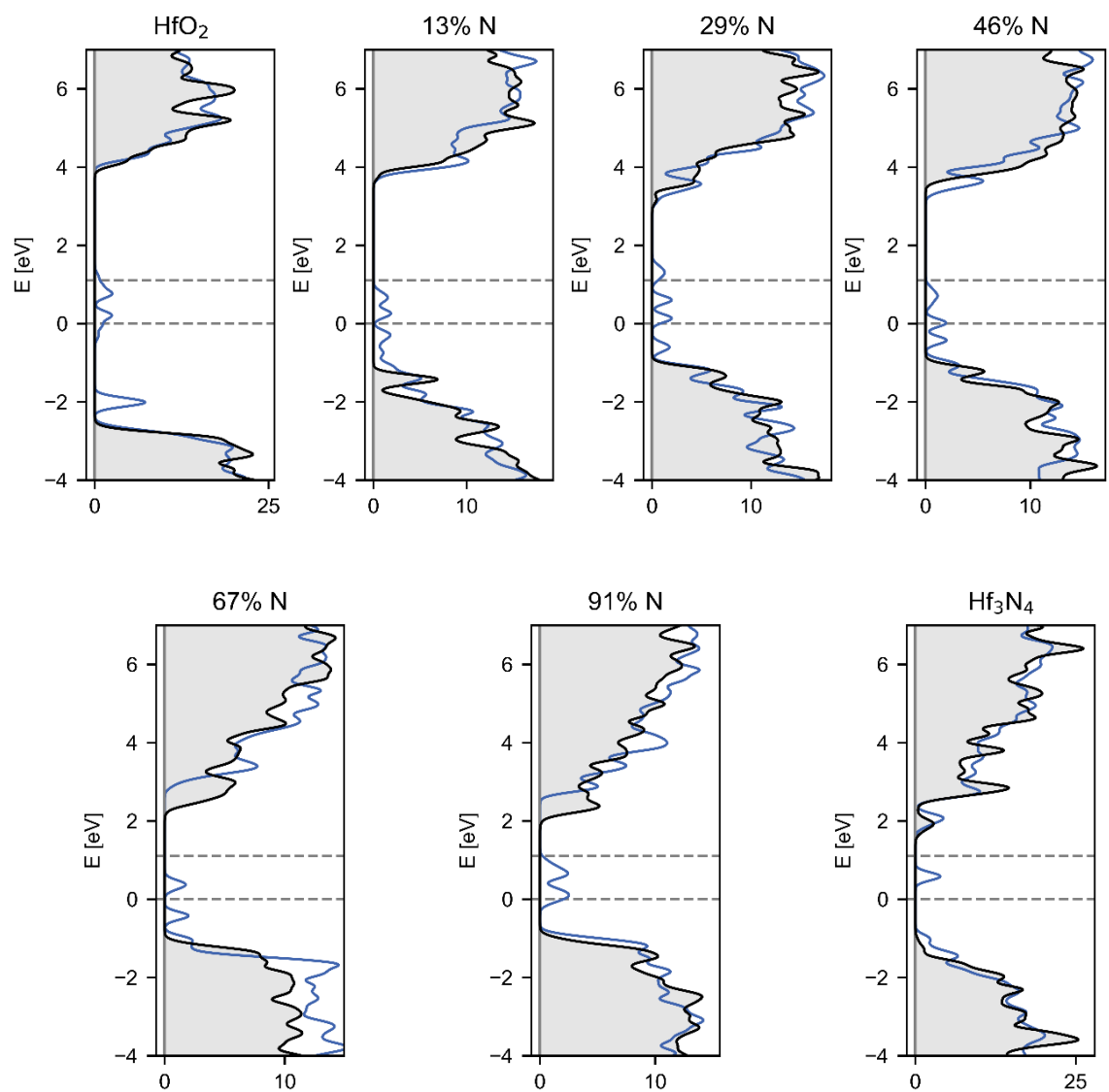

**Figure S11.** Density of states plots for N interstitial defects in a range of $HfO_xN_y$ compositions.

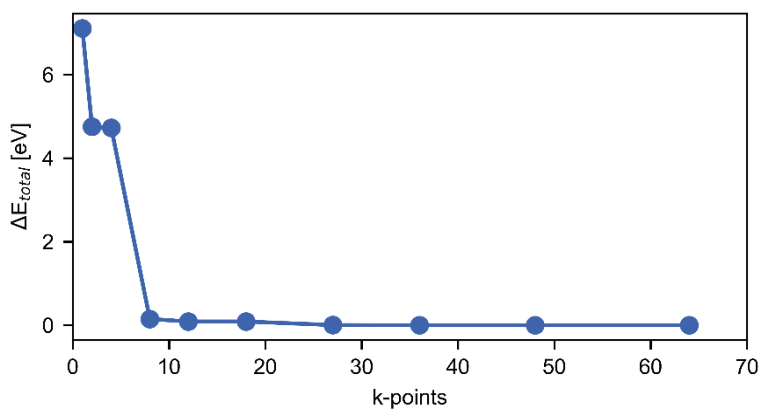

**Figure S12.** Convergence of total energy for the orthorhombic $HfO_2$ unit cell with increasing k-points.

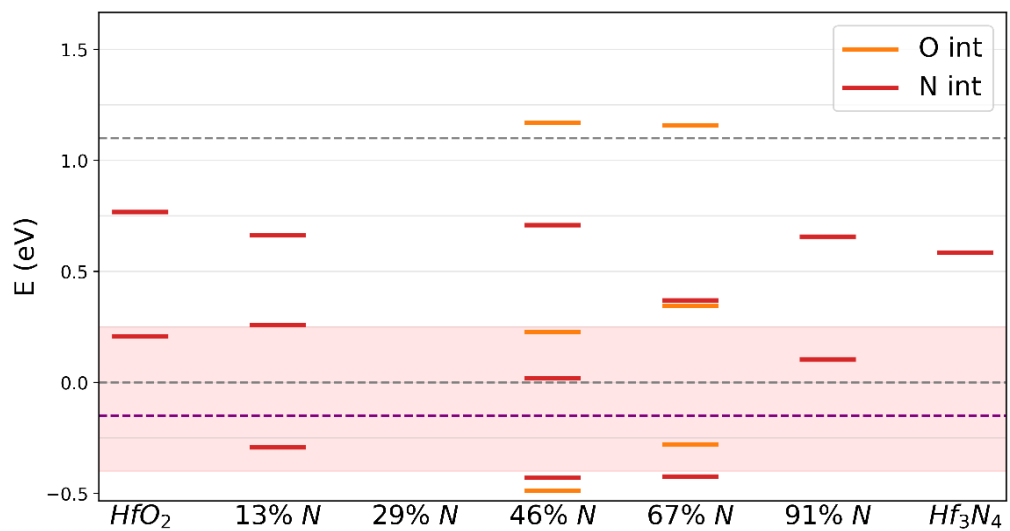

**Figure S13.** Summary of defect energy levels caused by O interstitials (orange) and N interstitials (red). The Si valence band maximum (VBM) and conduction band minimum (CBM) are marked with dotted grey lines, and the dotted purple line represents the minimum energy a defect state could have such that $E_\pm \leq E_{T,Tc}$. The red shaded box marks the energy region for potential hole traps. The energy axis is shifted such that the Si VBM is at 0 energy.

**Table 1.** Formation energies for N interstitial defects.

| Defect formation energy [eV] | $HfO_2$ | 13% N | 29% N | 46% N | 67% N | 91% N | $Hf_3N_4$ |
|---|---|---|---|---|---|---|---|
| N rich | 5.64 | 4.29 | 6.06 | 4.34 | 5.91 | 2.42 | 3.35 |
| N intermediate | 7.55 | 6.20 | 7.97 | 6.25 | 7.82 | 4.33 | 5.26 |
| N poor | | 9.45 | 8.11 | 9.87 | 8.16 | 9.73 | 6.24 | 7.17 |

**Table 2.** Formation energies for N vacancy defects.

| Defect formation energy [eV] | 13% N | 29% N | 46% N | 67% N | 91% N | $Hf_3N_4$ |
|---|---|---|---|---|---|---|
| N rich | 3.60 | 2.37 | 2.62 | 2.44 | 0.90 | 3.29 |
| N intermediate | 1.69 | 0.47 | 0.71 | 0.53 | -1.01 | 1.39 |
| N poor | -0.22 | -1.44 | -1.20 | -1.38 | -2.92 | -0.53 |

**Table 3.** Formation energies for O interstitial defects.

| Defect formation energy [eV] | $HfO_2$ | 13% N | 29% N | 46% N | 67% N | 91% N | $Hf_3N_4$ |
|---|---|---|---|---|---|---|---|
| O rich | 2.11 | 0.67 | -0.07 | -0.95 | 1.47 | -1.48 | 0.41 |
| O intermediate | 5.00 | 3.65 | 2.92 | 2.03 | 4.45 | 1.51 | 3.39 |
| O poor | 7.89 | 6.63 | 5.90 | 5.02 | 7.43 | 4.49 | 6.37 |

**Table 4.** Formation energies for O vacancy defects.

| Defect formation energy [eV] | $HfO_2$ | 13% N | 29% N | 46% N | 67% N | 91% N |
|---|---|---|---|---|---|---|
| O rich | 6.37 | 5.62 | 4.96 | 6.57 | 5.88 | 4.09 |
| O intermediate | 3.39 | 2.55 | 1.31 | 3.48 | 2.74 | 1.10 |
| O poor | 0.40 | -0.53 | -2.33 | 0.39 | -0.39 | -1.88 |

**Table 5.** Defect concentration, assuming O poor, N intermediate limits, and T = 423 K.

| Concentration [cm$^{-3}$] | $HfO_2$ | 13% N | 29% N | 46% N | 67% N | 91% N | $Hf_3N_4$ |
|---|---|---|---|---|---|---|---|
| O vacancy | $4.17 \times 10^{16}$ | $1.12 \times 10^{28}$ | $1.98 \times 10^{50}$ | $6.46 \times 10^{16}$ | $2.59 \times 10^{26}$ | $5.72 \times 10^{24}$ | N/A |
| N vacancy | N/A | $5.96 \times 10^{0}$ | $7.05 \times 10^{15}$ | $7.04 \times 10^{12}$ | $1.21 \times 10^{15}$ | $9.91 \times 10^{33}$ | $2.74 \times 10^{4}$ |
| O interstitial | $2.17 \times 10^{-76}$ | $6.95 \times 10^{-61}$ | $7.75 \times 10^{-52}$ | $5.93 \times 10^{-41}$ | $9.42 \times 10^{-71}$ | $1.67 \times 10^{-34}$ | $1.11 \times 10^{-57}$ |
| N interstitial | $4.06 \times 10^{-72}$ | $1.45 \times 10^{-55}$ | $2.69 \times 10^{-77}$ | $3.59 \times 10^{-56}$ | $1.77 \times 10^{-75}$ | $1.83 \times 10^{-32}$ | $4.15 \times 10^{-34}$ |